\documentclass[12pt]{article} 
\usepackage[dvips]{graphicx} 

\begin{document}

\title{Flavon exchange effects in models with abelian flavor symmetry}
\author{Ilja Dorsner and S.M. Barr\\
Bartol Research Institute\\
University of Delaware\\ 
Newark, DE 19716}

\maketitle

\begin{abstract}

In models with abelian flavor symmetry the small mixing angles and
mass ratios of quarks and leptons are typically given by powers of
small parameters characterizing the spontaneous breaking of flavor
symmetry by "flavon" fields. If the scale of the breaking of flavor
symmetry is near the weak scale, flavon exchange can lead to interesting 
flavor-violating and CP violating effects. These are studied. It is found 
that $d_e$, $\mu \rightarrow e + \gamma$, and $\mu$-$e$ conversion on nuclei
can be near present limits. For significant range of parameters $\mu$-$e$ 
conversion can be the most sensitive way to look for such effects.

\end{abstract}

\newpage

\section{Introduction}

Flavor symmetry was first proposed to explain the structure of
the quark and lepton mass spectrum and the CKM mixing of the quarks
\cite{Froggatt:1978nt,Berezhiani:hm}.
More recently these ideas have been extended to account for the
observed patterns of neutrino masses and mixings 
\cite{Berezhiani:1996bv}. 
In the context of supersymmetry, flavor symmetry has been invoked to solve the
problem of flavor changing neutral currents, i.e. ``the SUSY Flavor Problem"
\cite{Nir:1993mx,Ibanez:ig}.

A wide assortment of flavor symmetries has been suggested. In particular,
models based on both non-abelian and abelian symmetries have been
constructed. One virtue of non-abelian symmetries is that they can
lead to degenerate masses, which have various theoretical uses.
For example, one solution to
the SUSY flavor problem is to posit a near degeneracy of the 
squark/slepton masses of the first
two families. For another example, large neutrino mixing angles
can be obtained by positing nearly degenerate neutrino masses. 
However, in this paper we shall be interested in abelian flavor symmetries. 

The main use to which abelian symmetries have been put is to explain
the hierarchical patterns of quark and lepton masses and mixings.
It has been observed that all of the interfamily mass ratios and mixing
angles can be written as powers of one or two small parameters.
For example, the quark mass ratios and mixings can be written 
$V_{us} \sim \epsilon$, $V_{cb} \sim \epsilon^2$, $V_{ub} \sim \epsilon^3$,
$m_d/m_s \sim \epsilon^2$, $m_s/m_b \sim 0.4 \epsilon^2$, and so on,
where $\epsilon \sim \lambda \sim 0.2$, the Wolfenstein parameter \cite{Wolfenstein:1983yz} .
This has suggested to many theorists the idea that there is a weakly broken
abelian symmetry which distinguishes
fermions that are of the same type but of different families.
Suppose, for instance, that there is a $U(1)_F$ flavor symmetry, under which
the Standard Model Higgs has charge zero, the fermions $\psi^c_i$ and 
$\psi_j$ have charges $\bar{q_i}$ and $q_j$, and  a ``flavon" field $S$  
has charge $-1$. Then a Yukawa operator $\psi^c_i \psi_j H$ is
forbidden by the flavor symmetry, but the effective operator
$\psi^c_i \psi_j H (S/M_F)^{(\bar{q_i} + q_j)}$ is not. Such an effective operator
might arise from integrating out fields whose mass is of order
$M_F$, the ``flavor scale". If one assumes that the breaking of $U(1)_F$
is weak, in the sense that $\langle S \rangle/M_F = \epsilon \ll 1$,
then one has explained the fact that the effective mass parameter
of the term $\psi^c_i \psi_j$ is proportional to a 
power of the small quantity $\epsilon$.
This is the basic idea of Froggatt and Nielson \cite{Froggatt:1978nt}, 
which has inspired a 
very large number of models in the literature. 

Aside from having the potential to explain the hierarchies observed
among fermion masses and mixing angles, this idea can be used to
construct models in which the dangerous flavor-changing effects in
supersymmetric models are suppressed by ``flavor alignment" \cite{Nir:1993mx}. 
The idea here is that in the preferred basis defined by the abelian
flavor charge assignments the off-diagonal elements of both the
fermion mass matrices and the sfermion mass-squared matrices are suppressed
by powers of the small parameters which characterize flavor breaking
(i.e. parameters like $\epsilon$). Thus the fermion and sfermion mass
matrices are nearly ``aligned" by flavor symmetry. The angles 
expressing their misalignment are suppressed by powers of the small 
parameters. If this suppression is strong enough it would solve the
SUSY Flavor Problem.

In this paper we examine some of the possible consequences for 
phenomenology of the exchange of the ``flavon" fields themselves. 
A point that should be stressed from the outset is that there do not have 
to be such consequences at all. The reason is that the flavor scale
$M_F$ can be anything from the weak scale up to the Planck scale.
All that matters is that the ratio $\langle S \rangle/M_F$ 
of the flavon expectation value
(or values) to the flavor scale be somewhat smaller than 1.
If the flavor scale is much larger than the weak scale, then the
phenomenological effects of flavon exchange will be unobservably small.
In fact, many papers assume that the flavor scale is near the Planck scale, 
which is certainly a reasonable expectation. However, since we do not
know a priori what the flavor scale is, it is interesting to
investigate the phenomenology that would result from its being
near the weak scale, and in particular to ask how low the flavor 
scale could actually be given present limits on flavor-changing and 
CP-violating processes. We would also like to know in which processes 
flavon-exchange effects would be likely first to show up.

There are many ways that new flavor physics just
above the weak scale could affect low-energy phenomenology. For 
instance, if the abelian flavor group is local, the exchange of the
corresponding gauge bosons could cause flavor-changing neutral current
processes. We will assume that the flavor group is either global, or
breaks to a global symmetry at a sufficiently high scale that such
gauge-boson-exchange effects can be ignored. We are only interested
in this paper in the exchange of the flavon fields themselves.

There are many models with abelian flavor symmetry, and the number
of parameters in such models can be large. What we shall do, therefore,
is write down an effective low-energy theory that has a managably small 
number of parameters and that has some of the typical features of models 
with abelian flavor symmetry. Studying this toy model will give some idea 
of the likely magnitude of various effects. We will then look at some 
variations of the model to see how they would change the conclusions.

\section{A simple effective theory of flavon physics}

The model we shall study has a single flavon field $S$ that is a singlet
under the Standard Model gauge group. The effective Yukawa couplings of the
quarks and leptons to $S$ and to the ordinary Standard Model Higgs field $H$,
after integrating out all the fields whose mass is of order the flavor scale 
$M_F$, is assumed to be

\begin{equation}
\label{lagrangian}
{\cal L}^{Yukawa} = - \hat{\lambda}^u_{ij} \epsilon^{ab} \overline{Q}^a_{Li}
H^{\dag b} u_{Rj} - \hat{\lambda}^d_{ij} \overline{Q}^a_{Li} H^a d_{Rj}
- \hat{\lambda}^l_{ij} \overline{L}^a_{Li} H^a l_{Rj} + h.c.,
\end{equation}

\noindent
where $i,j = 1,2,3$ are family indices, and $a,b = 1,2$ are $SU(2)_L$
indices. The $\hat{\lambda}$'s are given by the following expressions.

\begin{equation}
\label{lambda}
\hat{\lambda}^u = \left( \begin{array}{ccc}
h^u_{11} \hat{\epsilon}^4 & h^u_{12} \hat{\epsilon}^4
& h^u_{13} \hat{\epsilon}^4 \\ h^u_{21} \hat{\epsilon}^4 &
h^u_{22} \hat{\epsilon}^2 & h^u_{23} \hat{\epsilon}^2 \\
h^u_{31} \hat{\epsilon}^4 & h^u_{32} \hat{\epsilon}^2
& h^u_{33} \end{array} \right), \;\;\;
\hat{\lambda}^d = \left( \begin{array}{ccc}
h^d_{11} \hat{\epsilon}^6 & h^d_{12} \hat{\epsilon}^6
& h^d_{13} \hat{\epsilon}^6 \\ h^d_{21} \hat{\epsilon}^6 &
h^d_{22} \hat{\epsilon}^4 & h^d_{23} \hat{\epsilon}^4 \\
h^d_{31} \hat{\epsilon}^6 & h^d_{32} \hat{\epsilon}^4
& h^d_{33} \hat{\epsilon}^2 \end{array} \right).
\end{equation}

\noindent
The corresponding matrix for the charged leptons, $\hat{\lambda}^l$ 
is assumed to have the same form as $\hat{\lambda}^d$ with 
$h^d_{ij} \rightarrow h^l_{ij}$. In these expressions 
the $h_{ij}$ are all assumed to be of order unity, and the
hierarchy among various masses and mixing angles therefore comes 
from the powers of $\hat{\epsilon}^2$, which is defined to be

\begin{equation}
\label{epsilon}
\hat{\epsilon}^2 \equiv  S/M_F.
\end{equation}

\noindent
The particular structure given in Eq. (\ref{lambda}) is inspired by a model of 
Babu and Nandi \cite{bn}, 
which has the same powers of $\hat{\epsilon}$, but where
$\hat{\epsilon} = (H^{\dag} H)/M_F^2$ rather than $S/M_F$ as here. 
Their model is not a typical flavon model, therefore. However, the
pattern of powers of $\epsilon$ is quite typical of many abelian flavon 
models, and gives, as Babu and Nandi show (see below), an excellent fit
to quark and lepton masses and CKM angles. 
If we call the vacuum expectation value of the flavon field
$\langle S \rangle \equiv u$, then the small parameter that characterizes
flavor changing is $\epsilon^2 \equiv u/M_F$.

The Higgs potential is assumed to have the form

\begin{equation}
\begin{array}{ccl} 
V(H,S) & = & \lambda(H^{\dag} H)^2 - \mu^2(H^{\dag} H) + \lambda_S(S^* S)^2
- \mu^2_S(S^* S) \\ & & \\ & + & \lambda'(H^{\dag} H S^* S) 
-\frac{1}{2} (\delta m^2 S^2 + h.c.). \end{array}
\end{equation}

\noindent
The last term has been put in to give a soft breaking of the 
global $U(1)_F$ under which $S \rightarrow \mathrm{e}^{\mathrm{i} \theta} S$, and thus
to give mass to the pseudoscalar part of $S$. (This global $U(1)$
may ultimately come from a local flavor symmetry that is broken
at a higher scale.) The parameter $\delta m^2$ is the only one in the
Higgs potential that can have a phase. However, one can absorb this
by a phase rotation of $S$. Having done so, the VEV of $S$ is a
real quantity. Minimizing this potential gives 

\begin{equation}
\label{scalars}
\begin{array}{ccl} 
S & = & u + \frac{1}{\sqrt{2}} s_1 + \frac{\mathrm{i}}{\sqrt{2}} s_2, \\
& & \\
H & = & \left( \begin{array}{c} 0 \\ v + \frac{1}{\sqrt{2}} h \end{array}
\right),
\end{array}
\end{equation}

\noindent
where

\begin{equation}
\begin{array}{ccl}
v^2 & = & [2 \lambda_S \mu^2 - \lambda'
(\mu^2_S + \delta m^2)]/(4\lambda \lambda_S - \lambda^{\prime 2}), \\
& & \\
u^2 & = & [2 \lambda (\mu_S^2 + \delta m^2) - \lambda'
\mu^2]/(4\lambda \lambda_S - \lambda^{\prime 2}),
\end{array}
\end{equation}

\noindent
with $v \simeq 174$ GeV, and $\langle s_1 \rangle = \langle s_2 \rangle = \langle h \rangle = 0$.
From Eqs. (\ref{epsilon}) and (\ref{scalars}), we can write

\begin{equation}
\label{eps}
\hat{\epsilon}^2 = \epsilon^2 [ 1 +  (s_1 + \mathrm{i} s_2)/(\sqrt{2} u) ].
\end{equation}

\noindent
Consequently, the couplings of $s_1$ and $s_2$ to the quarks and leptons are
obtained by taking in Eq. (\ref{lagrangian})

\begin{equation}
\label{coupling}
\hat{\lambda}^f_{ij} H = \lambda^f_{ij}  \hat{\epsilon}^{2n^{f}_{ij}} \left(v + \frac{h}{\sqrt{2}}\right) 
\cong  m^f_{ij} \left[1 + \frac{n^f_{ij} (s_1+ \mathrm{i} s_2)}{\sqrt{2}u} + \frac{h}{\sqrt{2}v}\right],
\end{equation}

\noindent
where $f = u,d,l$, and 
where $n^{f}_{ij}$ is the power of $\hat{\epsilon}^2$ that appears in 
$\hat{\lambda}^f_{ij}$. 
It turns out that for the interesting phenomenology one can ignore the
terms higher than linear in the fields $s_1$ and $s_2$ in Eq. (\ref{coupling}).
Note that the coupling of $h$ to
the quarks and leptons will be made real and diagonal when
the mass matrices $m^f_{ij}$ are,
but that the coupling of the flavon fields $s_1$ and $s_2$ will not be made real and
diagonal because of
the extra factor of $n^f_{ij}$. This is what will give the flavor-changing
and CP-violating effects that we shall be interested in.
We see also that $s_2$ couples in the same way to
quarks and leptons as $s_1$ does so but with a relative phase of $\mathrm{i}$. 
This factor of $\mathrm{i}$ comes in squared in $s_2$ exchange and so does not
lead to CP-violating effects to the order we are interested in.

Let us now look at how many parameters the model has. First, there are
the large number of parameters that we have called $h^f_{ij}$.  
Because there are so many, there is no hope of making any sharp predictions.
However, if we confine our ambition to making order of magnitude estimates
of effects, then we can (for the most part) ignore the $h^f_{ij}$, since
they are assumed all to be of order unity. This leaves the six
parameters in the Higgs potential $(\lambda, \mu^2, \lambda_S, \mu^2_S,
\lambda', \delta m^2)$, and the flavor scale $M_F$. These parameters
can be traded for $v$, $m_h$, $u$, $m_{s_1}$, $\sin \phi$, $m_{s_2}$,
and $M_F$. The VEV $v$ is known precisely; the mass of the ordinary higgs
$m_h$ is known approximately; and the parameter $M_F$ is determined
by the relation $\epsilon^2 \equiv u/M_F$. (The value of $\epsilon^2$ is
known approximately from the values of the quark and lepton mass ratios and
the CKM angles.) Consequently, one is left
with four free parameters: the masses of the scalar flavon $m_{s_1}$ and
the pseudoscalar flavon $m_{s_2}$,
the VEV $u$ of the flavon (which, as we have seen, controls the strength
of the flavon couplings to matter), and the parameter $\sin \phi$ 
that describes the mixing between the ordinary Higgs and the scalar flavon. 
This mixing is described by the mass matrix

\begin{equation}
\frac{1}{2} ( h\;\; s_1\;\; s_2) \left( \begin{array}{ccc}
4 \lambda v^2 & 2 \lambda' v u & 0 \\
2 \lambda' v u & 4 \lambda_S u^2 & 0 \\
0 & 0 & 2 \delta m^2 \end{array} \right) \left( \begin{array}{c}
h \\ s_1 \\ s_2 \end{array} \right).
\end{equation}

\noindent
so that $\tan 2 \phi = (\lambda' v u)/(\lambda_s u^2 - \lambda v^2)$.
We will call the mass eigenstates $h' = \cos \phi \,h -
\sin \phi \,s_1$, and $s' = \sin \phi \,h + \cos \phi \,s_1$, and
their masses $m_{h'}$, and $m_{s'}$, respectively.

Turning to the diagonalization of the quark and lepton mass matrices, one
finds that

\begin{eqnarray}
(m_u, m_c, m_t) & \cong & \left( |h_{11}^u - h_{12}^u h_{21}^u/h_{22}^u| 
\epsilon^6, |h_{22}^u| \epsilon^2, |h_{33}^u| \right) v, \nonumber \\
(m_d, m_s, m_b) & \cong & \left( |h_{11}^d| \epsilon^6, |h_{22}^d| \epsilon^4,
|h_{33}^d| \epsilon^2 \right) v, \\
(m_e, m_{\mu}, m_{\tau} ) & \cong & \left( |h_{11}^l| \epsilon^6,
|h_{22}^l| \epsilon^4, |h_{33}^l| \epsilon^2 \right)v, \nonumber
\end{eqnarray}

\noindent
and

\parbox{9cm}{
\begin{eqnarray*}
\left| V_{us} \right| & \cong & \left| \frac{h_{12}^d}{h_{22}^d} - \frac{h_{12}^u}{h_{22}^u}
\right| \epsilon^2,\\
\left| V_{cb} \right| & \cong & \left| \frac{h_{23}^d}{h_{33}^d} - \frac{h_{23}^u}{h_{33}^u}
\right| \epsilon^2,\\
\left| V_{ub} \right| & \cong & \left| \frac{h_{13}^d}{h_{33}^d} - \frac{h_{13}^u}{h_{33}^u}
- \frac{h_{12}^u h_{23}^d}{h_{22}^u h_{33}^d}
+ \frac{h_{12}^u h_{23}^d}{h_{22}^u h_{33}^d} \right| \epsilon^4.
\end{eqnarray*}}
\hfill
\parbox{2cm}{
\begin{equation}
\label{CKM}
\end{equation}}

Babu and Nandi \cite{bn} showed that this gives a reasonable fit to the data.
They took the data to be 
$m_u \mathrm{(1 GeV)} = 5.1$ MeV, $m_d \mathrm{(1 GeV)} = 8.9$ MeV, $m_s \mathrm{(1 GeV)} = 
175$ MeV, $m_c(m_c) = 1.27$ GeV, $m_b(m_b) = 4.25$ GeV, $m_t^{phys} =
175$ GeV, $m_{\tau} = 1.78$ GeV, $m_{\mu} = 105.6$ MeV, and $m_e = 511$ keV.
Extrapolating, using the 3-loop QCD and one-loop QED beta
functions, with $\alpha_s (M_Z) = 0.118$, they obtained
the running masses in GeV evaluated at $m_t$:
$m_t \simeq 166$, $m_c \simeq 0.6$, $m_u \simeq 0.0022$, 
$m_b \simeq 2.78$, $m_s \simeq 0.075$, $m_d \simeq 0.0038$,
$m_{\tau} \simeq 1.75$, $m_{\mu} \simeq 0.104$, and $m_e \simeq 
0.0005$. These are well fit by $\epsilon^2 \simeq (1/6.5)^2 \cong
0.024$, if one takes $|h_{11}^u - h_{12}^u h_{21}^u/h_{22}^u|
\simeq 0.95$, $|h_{22}^u| \simeq 0.14$, $|h_{33}^u| \simeq 0.96$,
$|h_{11}^d| \simeq 1.65$, $|h_{22}^d| \simeq 0.77$, $|h_{33}^d|
\simeq 0.68$, $|h_{11}^l| \simeq 0.21$, $|h_{22}^l| \simeq 1.06$,
and $|h_{33}^l| \simeq 0.42$.

Note that with the exception of $h_{22}^u$ and $h_{11}^l$ all these
are of order unity. And as emphasized in \cite{bn} the smallness of
$h_{22}^u$ actually helps account for the values of $V_{us}$ and
$V_{ub}$. From Eq. (\ref{CKM}) one sees that with $h_{22}^u \simeq 1/7$,
these mixings come out to be  $V_{us} \sim 7 \epsilon^2 \sim 
0.2$, and $V_{ub} \sim 7 \epsilon^4 \sim 3 \times 10^{-3}$.

As mentioned, in the basis where the mass matrices of the quarks and 
leptons are diagonal and real, the couplings of $s_1$ and $s_2$ remain with
off-diagonal and complex elements, due to the extra factors of $n^{f}_{ij}$
in Eq. (\ref{coupling}). However, it is interesting that the flavor-diagonal
couplings of $s_1$ are, in fact, real to leading order in the small
parameter $\epsilon^2$. That is, the imaginary part of these diagonal
couplings is of order $\epsilon^2 \simeq 0.02$ times the real part.
This is significant for the lepton and quark electric dipole moments,
as we shall see. The reason that the diagonal couplings of $s_1$ are
real to leading order can be seen by looking at a simple two-by-two
example:

\begin{equation}
Y_h = \left( \begin{array}{cc}
h_{11} \epsilon^{2 n_{11}} & h_{12} \epsilon^{2 n_{12}} \\
h_{21} \epsilon^{2 n_{21}} & h_{22} \epsilon^{2 n_{22}} \end{array}
\right), \;\;\;
Y_{s_1} = \left( \begin{array}{cc}
h_{11} n_{11} \epsilon^{2 n_{11}} & h_{12} n_{12} \epsilon^{2 n_{12}} \\
h_{21} n_{21} \epsilon^{2 n_{21}} & h_{22} n_{22}
\epsilon^{2 n_{22}} \end{array}
\right) \frac{v}{u}.
\end{equation}

\noindent
In the basis where $Y_h$ is diagonal and real,
which we shall denote by primes, 

\begin{eqnarray}
\label{Yprime}
(Y'_h)_{11} & \cong & \left[ h_{11} \epsilon^{2 n_{11}}
- \frac{h_{12} h_{21}}{ h_{22}} 
\epsilon^{2(n_{12} + n_{21} - n_{22})} \right] \mathrm{e}^{\mathrm{i} \alpha},
\nonumber\\ & & \\
(Y'_{s_1})_{11} & \cong & \left[ h_{11} n_{11} \epsilon^{2 n_{11}}
- \frac{h_{12} h_{21}}{ h_{22}} (n_{12} + n_{21} - n_{22}) \epsilon^{2
(n_{12} + n_{21} - n_{22})} \right] \mathrm{e}^{\mathrm{i} \alpha}\frac{v}{u}.\nonumber
\end{eqnarray}

\noindent
The factor of $\mathrm{e}^{\mathrm{i} \alpha}$ is the phase rotation required to make
$(Y'_h)_{11}$ real.
(In the same basis, the matrix $Y'_{s_1}$ is easily seen to be non-diagonal:
$|(Y'_{s_1})_{12}| \cong |h_{12}(n_{12} - n_{22}) \epsilon^{2 n_{12}}|(v/u)$,
and $|(Y'_{s_1})_{21}| \cong |h_{21}(n_{21} - n_{22}) \epsilon^{2 n_{21}}|
(v/u)$.)
In Eq. (\ref{Yprime}) one sees two terms in the expression for $(Y'_{h})_{11}$.
There are two cases to consider: either these two terms are 
of the same order in $\epsilon^2$, 
or one is higher order in $\epsilon^2$ than the other.
If they are the same order, then $n_{12} + n_{21} - n_{22} = n_{11}$, which
means that $(Y'_{s_1})_{11} = n_{11} (Y'_h)_{11} (v/u)$, a real quantity,
to leading order in $\epsilon^2$. 
If, on the other hand, one term in $(Y'_h)_{11}$ is of lower
order in $\epsilon^2$ than the other and dominates,
then the corresponding term dominates in $(Y'_{s_1})_{11}$. Consequently,
to leading order in $\epsilon^2$, one has again that $(Y'_{s_1})_{11}$
is just an integer times $(Y'_h)_{11} (v/u)$ and therefore real.

This conclusion generalizes to more complicated situations. It is true
for $N$-by-$N$ matrices. It is also true if
there are several abelian flavon fields giving several $\epsilon$
parameters, as long as contributions to diagonal Yukawa couplings that are
of different orders in the small parameters are not accidentally
numerically comparable.

\section{Flavor-changing and CP-violating \\processes}

We are now ready to discuss various flavor-changing and CP-violating
processes. The ones that shall be of chief interest are
$\Delta m_K^2$ and $\epsilon_K$ in the neutral Kaon system, the
electric dipole moment of the electron $d_e$, 
the decay $\mu \rightarrow e + \gamma$, and $\mu$-$e$ 
conversion on nuclei $\mu + N \rightarrow e + N$. 
It is straightforward to calculate the contributions to these effects
coming from flavon exchange in our toy model. 

The relevant couplings for flavor-changing and CP-violating processes, 
in  the physical basis of fermions and bosons, can be parametrized as  

\begin{equation} 
{\cal L} = -\frac{\sqrt{m_i m_j}}{v} \bar{f_i}(\Delta_{ij}^{a\,L}P_L
+\Delta_{ij}^{a\,R}P_R) f_j H_a
+g\, m_W\cos\varphi_a W^+ W^- H_a+\cdots, 
\end{equation}  

\noindent 
where $a=h', s'$, $P_{L,R}=(1\mp \gamma_5)/2$, and where indices $i$ and $j$ run over 
all quarks and charged leptons. 
We observe that due to the scalar nature of $h'$ and $s'$, to the leading order in 
$\epsilon^2$, $\Delta_{ii}^{a\,L} \equiv \Delta_{ii}^{a\,R*}= \Delta_{ii}^{a\,R} 
\equiv \Delta_{ii}^{a}$ is real for all $i$'s. (See the discussion after Eq. (\ref{Yprime}).) 
Acting on Yukawa coupling matrices with a set of bi-unitary transformations that 
brings fermion mass matrices into diagonal form, 
and simultaneously diagonalizing the Higgs sector one finds that   

\parbox{5.5cm}{ 
\begin{eqnarray*} 
\Delta_{ee}^{h'\, L} & = & 4 \chi^2 \epsilon^2 \frac{h_{12}^l h_{21}^l}{\sqrt{2}} \sin\phi \frac{v}{u},\\ 
\Delta_{e \mu}^{h'\, L} & = & - \chi \epsilon \frac{h_{12}^l}{\sqrt{2}} \sin\phi \frac{v}{u}, 
\end{eqnarray*}} 
\hfill 
\parbox{5.5cm}{ 
\begin{eqnarray*} 
\Delta_{ee}^{h'\, R} & = & 4 \chi^2 \epsilon^2 \frac{h_{12}^{l*} h_{21}^{l*}}{\sqrt{2}} \sin\phi \frac{v}{u},\\ 
\Delta_{e \mu}^{h'\, R} & = & - \chi \epsilon \frac{h_{21}^{l*}}{\sqrt{2}} \sin\phi \frac{v}{u}, 
\end{eqnarray*}} 
\hfill 
\parbox{1.5cm}{ 
\begin{equation} 
\end{equation}}  

\noindent 
and  

\begin{equation} 
\cos\varphi_{h'}=\cos\phi,\qquad \qquad \cos\varphi_{s'}=\sin\phi,
\end{equation}  

\noindent 
where we have omitted a term in $\Delta_{ee}$ which is real and  
leading order in $\epsilon^2$, and introduced 
$\chi = (|h_{11}^l||h_{22}^l|)^{-1/2}$. 
The coefficients $\Delta_{ij}^{s'\, L,R}$ 
are obtained from $\Delta_{ij}^{h'\, L,R}$ by making the transformation 
$\cos \phi \rightarrow \sin \phi$, and $\sin \phi \rightarrow -\cos \phi$.  

\begin{figure} 
\begin{center} 
\includegraphics[width=4in]{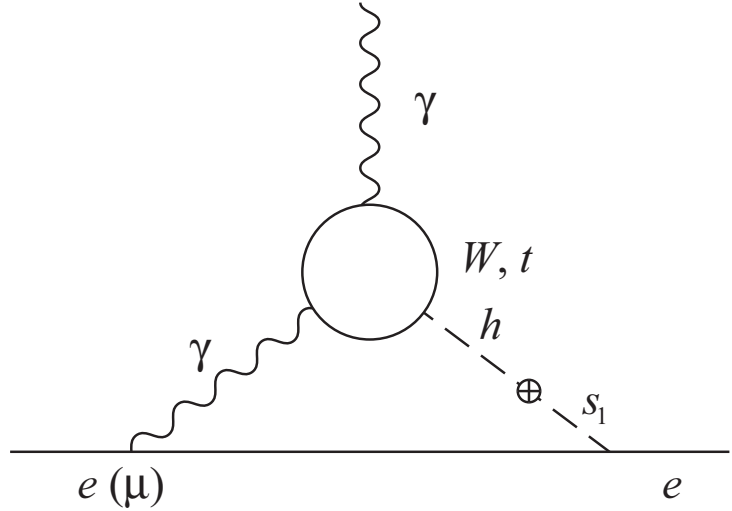} 
\end{center} 
\caption{\label{edm} A two-loop Feynman diagram for electron electric 
dipole moment ($\mu \rightarrow e+\gamma$).} 
\end{figure}

The electric dipole moment of the electron ($d_e$) comes from the familiar
type of two-loop graph \cite{Barr:vd} shown in Fig. \ref{edm}. 
In terms of the original
fields $s_1$, $s_2$ and $h$ coming from $S$ and $H$, rather than the mass 
eigenstates, one sees that the field that couples to the 
$W$ or $t$ loop must be $h$.
This can be seen as follows. The $s_i$ have no coupling to $t$ at
the leading order in $\epsilon^2$, since the $t$ mass comes from
order $(S/M_F)^0$. i.e. $n^u_{33} = 0$. (See Eqs. (\ref{lambda}) and
(\ref{coupling}).) The $s_i$ also have no coupling to the $W^{\pm}$ 
since $S$ does not
participate in breaking $SU(2)_L \times U(1)_Y$. (If there were two
Higgs doublets in the model, then the heavy loop could be a charged
Higgs, in which case the field coupling to it in Fig. \ref{edm} could be an
$s_1$.) However, the field coupling to the electron line 
must be either $s_1$ or $s_2$ in order to obtain a CP-violating phase,
since the couplings of $h$ are real and flavor diagonal in the physical
basis of the leptons. However, the $s_2$, while it can give a CP-violating
phase, does not mix with the $h$ and therefore would not be able to
attach to the $W$ or $t$ loop. The scalar line in the two-loop graph for $d_e$
is thus $s_1$ where it attaches to the electron, and $h$ where it attaches
to the $W$ or $t$ loop. Consequently,
the electron edm is proportional to the mixing $\sin \phi \cos \phi$. 
A significant point
about the $d_e$ diagram, which has already been alluded to, is that
while the $s_1$ coupling to the electron has a CP-violating phase, that
phase brings in an extra suppression of order $\epsilon^2$ (see Eq. (15)). 
The electric dipole moment of a charged lepton is given by 

\begin{equation} 
d_i = \frac{e\, G_F \alpha }{8 \sqrt{2} \pi^3 } 
m_i\, \mbox{Im}[A_{ii}^{L}- A_{ii}^{R}], 
\end{equation} 
 
\noindent 
where the dominant, reduced amplitude \cite{Chang:kw}, comes from $W$ loop 
and reads  

\begin{equation} \label{amplitude} 
A_{ij}^{L,R}=-\sum_{a} \cos \varphi_a  
\Delta_{ij}^{a\, L,R} 
\left [3 f\left(\frac{m^2_W}{m^2_{a}}\right) + 
\frac{23}{4} g\left(\frac{m^2_W}{m^2_{a}}\right) 
+ \frac{3}{4} h\left(\frac{m^2_W}{m^2_{a}}\right)\right]. 
\end{equation}  

\begin{figure} 
\begin{center} 
\includegraphics[width=4in]{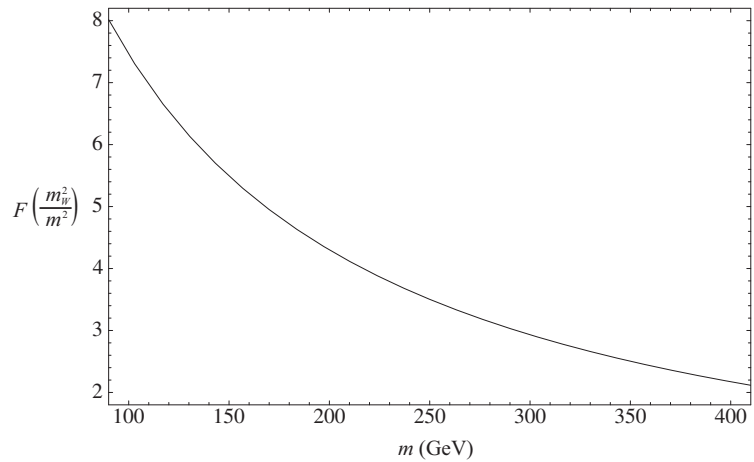} 
\caption{\label{F} Plot of $F(m_W^2/m^2) \equiv 3 f(m_W^2/m^2) + \frac{23}{4} g(m_W^2/m^2) 
+ \frac{3}{4} h(m_W^2/m^2)$ as a function of scalar mass $m$. The functions $f$, $g$, and $h$ 
are as defined in \cite{Chang:kw} in Eqs. (10), (11), and (15), respectively.} 
\end{center} 
\end{figure}

\noindent 
Setting $h_{12}^l=h_{21}^l=\mathrm{e}^{-\mathrm{i} \pi /4}$ in Eq. (15), the electron 
edm comes out to be

\begin{equation}
\label{edmlimit}
d_e = (1.5 \times 10^{-27} \;e\,{\rm cm}) \sin \phi \cos \phi \left( \frac{v}{u}
\right) \left[ F \left( \frac{m^2_W}{m^2_{h'}} \right)
- F \left( \frac{m^2_W}{m^2_{s'}} \right) \right].
\end{equation}

\noindent
where $F(z) \equiv 3 f(z) + \frac{23}{4} g(z) + \frac{3}{4} h(z)$,
and the functions $f$, $g$, and $h$ are as defined in \cite{Chang:kw} in Eqs. (10), (11), 
and (15), respectively. The function $F(z)$ is plotted in Fig. \ref{F}. Using the
experimental value $d_e = 0.18 \times 10^{-26} \;e\,{\rm cm}$ \cite{Groom:in} gives the following limit

\begin{equation}
\label{de}
\sin \phi \cos \phi \left( \frac{v}{u} \right) 
\left[ F \left( \frac{m^2_W}{m^2_{h'}} \right)
- F \left( \frac{m^2_W}{m^2_{s'}} \right) \right] \leq 1.2.
\end{equation}

The diagram for 
$\mu \rightarrow e + \gamma$ is of the same two-loop type as the 
electron edm diagram, except that one of the external leptons
is a $\mu$ rather than an $e$. 
As in the $d_e$ case, the scalar which couples to the 
lepton line must be $s_1$ (here because it involves flavor-changing), 
while the scalar that couples to the $W^{\pm}$ or
$t$ loop must be $h$. Thus the amplitude here is also proportional
to $\sin \phi \cos \phi$. The branching ratio for the process  
$l_j \rightarrow l_i + \gamma$ 
is given by  

\begin{equation} 
B(l_j \rightarrow l_i + \gamma) = \frac{3}{4} \left(\frac{\alpha }{\pi }
\right)^3 \frac{m_i}{m_j} 
\left(\frac{1}{2} |A_{ij}^L|^2 + \frac{1}{2} |A_{ij}^R|^2 \right). 
\end{equation}  

\noindent 
Again setting $h_{12}^l=h_{21}^l=\mathrm{e}^{-\mathrm{i} \pi /4}$,
and imposing the experimental limit $B(\mu \rightarrow e + \gamma)\le 1.2 
\times 10^{-11}$ 
\cite{Brooks:1999pu}, one obtains 

\begin{equation}
\label{muegamma}
\sin \phi \cos \phi \left( \frac{v}{u} \right) 
\left[ F \left( \frac{m^2_W}{m^2_{h'}} \right)
- F \left( \frac{m^2_W}{m^2_{s'}} \right) \right] \leq 2.2.
\end{equation}

\begin{figure} 
\begin{center} 
\includegraphics[width=4in]{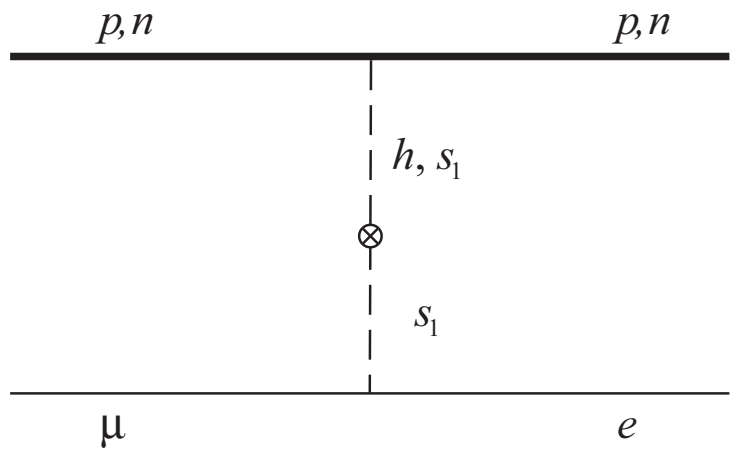} 
\caption{\label{mutoe} Tree level scalar exchange Feynman diagram for 
$\mu$-$e$ conversion on nuclei.} 
\end{center} 
\end{figure}

The diagram relevant for $\mu$-$e$ conversion on nuclei is Fig. \ref{mutoe}.
The field that couples to the lepton line must be $s_1$ or $s_2$, but the
field that couples at the quark line may be $h$, $s_1$, or $s_2$.
It is well known that the contributions of the pseudoscalar exchange to the 
coherent $\mu$-$e$ conversion
on nuclei can be neglected \cite{Kuno:1999jp} and will be ignored in our calculations.
The contributions to the amplitude from diagrams where the scalar couples
to the lepton as $s_1$ but to the quark as $h$ go as 
$\sin \phi \cos \phi(1/m^2_{h'} - 1/m^2_{s'})$. Those in which
the scalar couples to both the lepton line and the quark line as $s_1$
go as $\cos^2 \phi (1/m^2_{s'}) + \sin^2 \phi (1/m^2_{h'})$. 
We shall see these expressions emerge in the formulas that appear below. 

The branching ratio of $\mu$-$e$ conversion $B(\mu^- + (A,Z) \rightarrow e^- + (A,Z))$, defined 
to be the ratio of decay widths $\Gamma (\mu^- + (A,Z) \rightarrow e^- + (A,Z))/
\Gamma (\mu^- + (A,Z) \rightarrow capture)$, can be found using the procedure outlined 
in \cite{Weinberg:cj,Marciano:cj}. 
We obtain
\begin{equation} 
B=2 G_F^2 m_e m_{\mu} \frac{\alpha^3 m_{\mu}^5 Z_{eff}^4}{\pi^2 Z \Gamma_{capt}} A^2 F(q^2)^2
\left[ \left| \sum_{a} \frac{\tilde{m}_N^a}{m_{a}^2} \Delta_{e \mu}^{a\,L} \right|^2 
      +\left| \sum_{a} \frac{\tilde{m}_N^a}{m_{a}^2} \Delta_{e \mu}^{a\,R} \right|^2 \right], 
\end{equation}  

\noindent 
where $F(q^2)$ is the nucleon form factor, $Z_{eff}$ is the effective atomic number, and where
$\tilde{m}_N^a$ contains the heavy quark effects in effective 
scalar-nucleon-nucleon coupling \cite{Ng:1993ey} and is given by 

\begin{equation} 
\tilde{m}_N^a=\langle N |\sum_{l=u,d,s} m_l \Delta_{ll}^{a} \bar{f_l} f_l+\sum_{h=t,b,c} m_h \Delta_{hh}^{a} 
\bar{f_h} f_h | N \rangle. 
\end{equation}  

\noindent
We derive the most general, model independent, expression  for $\tilde{m}_N^a$ 
using the approach of Shifman et. al. \cite{Shifman:zn},
and subsequent improvements of inclusion of strange and heavy quark contributions
discussed in \cite{Cheng:1988cz,Gasser:1990ce} as follows\footnote{Our general expression 
for $\tilde{m}_N^a$ reproduces Eq. (3) of Ref. \cite{Ng:1993ey} but yields 
an additional term in Eq. (20) where authors analyze MSSM model. The additional piece is 
$\sigma_{\pi N}(\cot \beta + \tan \beta )/2$.}

\begin{equation} 
\tilde{m}_N^a=(\sum_{h} \Delta_{hh}^{a})
\frac{2}{27}\left[m_N-\sigma_{\pi N} \left(1+\frac{y}{2}\frac{m_s}{\bar{m}}\right) \right] 
+\sigma_{\pi N}\left[\frac{\Delta_{uu}^{a}+\Delta_{dd}^{a}}{2}+\Delta_{ss}^{a}\frac{y}{2}\frac{m_s}{\bar{m}}\right], 
\end{equation}  

\noindent
where $h$ runs over heavy quarks ($t,b,c$), $y=2 \langle N | \bar{s} s | N \rangle / \langle N | \bar{u} u +
\bar{d} d| N \rangle$ is the strange content in the nucleon, 
$\sigma_{\pi N}$ is the pion-nucleon sigma term, 
$m_N$ is the nucleon mass, and $\bar{m}=(m_u+m_d)/2$. In our model, the diagonalization procedure in
quark sector, to the leading order in $\epsilon^2$, 
leads to

\begin{equation} 
\Delta_{ii}^{h'}=\left[\cos \phi - \kappa_{i} \frac{v}{u} \sin \phi \right]/\sqrt{2}, 
\end{equation}  

\noindent 
where $(\kappa_{u},\kappa_{c},\kappa_{t},\kappa_{d},\kappa_{s},\kappa_{b})=(3,1,0,3,2,1)$. 
Note that $\kappa_{i}$'s are the powers of $\hat{\epsilon}^2$ of the appropriate diagonal
elements that appear in $\hat{\lambda}^f_{ij}$ of Eq. (\ref{lambda}).

For $\mu$-$e$ conversion on $^{48}_{22}\mathrm{Ti}$, we set $Z_{eff}=17.6$, $F(q^2=-m_{\mu}^2)=0.54$, 
$\Gamma_{capt}=2.59 \times 10^6 \; \mathrm{s}^{-1}$ \cite{Suzuki:1987jf}, impose the experimental limit 
$B < 4.3 \times 10^{-12}$ \cite{Dohmen:mp}, take $\bar{m}=5$ MeV, and use the set 
$(y,\sigma_{\pi N})=(0.47,60 \mbox{MeV})$ \cite{Cheng:1988cz}, to obtain

\begin{equation}
\label{mueconversion}
\left( \frac{v}{u} \right) 
\left| \sin \phi \cos \phi \left( 
\frac{1}{m_{h'}^2} - \frac{1}{m_{s'}^2} \right) m 
- \left( \frac{v}{u} \right) \left( \frac{\sin^2 \phi}{m_{h'}^2}
+ \frac{\cos^2 \phi}{m_{s'}^2} \right) m' 
\right| \leq \frac{9 \times 10^{-5}}{{\rm GeV}},
\end{equation}

\noindent
where $m \simeq 350$ MeV, and $m' \simeq 500$ MeV. 

\begin{figure} 
\begin{center} 
\includegraphics[width=4in]{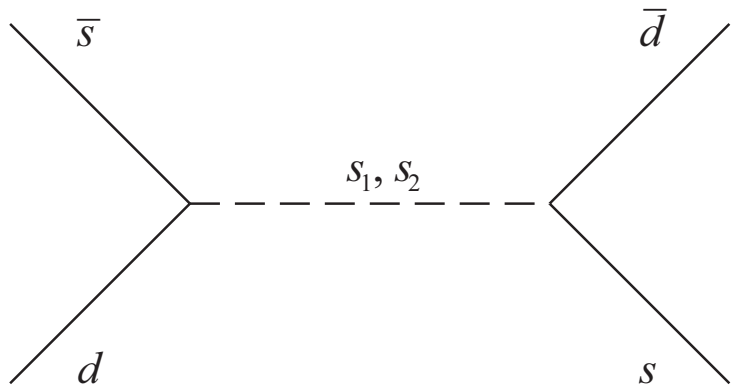} 
\caption{\label{KKbar} Tree level contribution to $K^0 - \bar{K}^0$ mixing.} 
\end{center} 
\end{figure}

The diagram relevant for the $\Delta S = 2$ processes is Fig. \ref{KKbar}.
Here, the field that couples at both quark lines must be $s_1$ or $s_2$.
Thus there are contributions that go as $\cos^2 \phi(1/m^2_{s'}) + 
\sin^2 \phi(1/m^2_{h'})$ and as $1/m^2_{s_2}$. 
Noting that $\Delta_{ds}^{L,R}$ is obtained from $\Delta_{e\mu}^{L,R}$ 
(Eq. (15)) by replacing $h_{ij}^l$ with $h_{ij}^d$, and using the vacuum saturation 
approximation for the hadronic element \cite{Atwood:1996vj}, we find a 
new contribution coming from the scalar exchange to be

\begin{eqnarray}
\label{deltaepsilon}
\epsilon_K^a & \simeq & \frac{C_K }{m^2_a}
\left\{ \left( \frac{1}{6}\frac{M_K^2}{(m_d+m_s)^2}+\frac{1}{6} \right)\mbox{Im} \left[ 
\left(\frac{h_{12}^{d*}+h_{21}^{d}}{\sqrt{2}} \right)^2 \right] \right. \nonumber\\
& - & \left. \left( \frac{11}{6}\frac{M_K^2}{(m_d+m_s)^2}+\frac{1}{6} \right)\mbox{Im} \left[
\left(\frac{h_{12}^{d*}-h_{21}^{d}}{\sqrt{2}} \right)^2 \right] \right\} (1-\cos^2 \varphi_a),
\end{eqnarray}

\noindent
while the exchange of pseudoscalar $s_2$, due to the extra factor of i, yields 

\begin{eqnarray}
\label{deltaepsilon1}
\epsilon_K^{s_2} & \simeq & \frac{C_K }{m^2_{s_2}}
\left\{ \left( \frac{1}{6}\frac{M_K^2}{(m_d+m_s)^2}+\frac{1}{6} \right)\mbox{Im} \left[ 
\left(\frac{h_{12}^{d*}-h_{21}^{d}}{\sqrt{2}} \right)^2 \right] \right. \nonumber\\
& - & \left. \left( \frac{11}{6}\frac{M_K^2}{(m_d+m_s)^2}+\frac{1}{6} \right)\mbox{Im} \left[
\left(\frac{h_{12}^{d*}+h_{21}^{d}}{\sqrt{2}} \right)^2 \right] \right\} \parbox{2.4cm} ,
\end{eqnarray}

\noindent
where we introduce 
\begin{equation}
\label{CK}
C_K=\frac{f^2_K M_K B_K \epsilon^{12}}{8 \sqrt{2}\, \Delta M_K}
\left(\frac{v}{u}\right)^2.
\end{equation}

\noindent
Using $B_K = 0.75$, $\Delta M_K \simeq 3.49 \times 10^{-12} 
$ MeV, $f_K \simeq 160$ MeV, $M_K \simeq 497.67$ MeV, $m_s = 175$ MeV, $m_d = 8.9$ MeV, 
$h_{12}^d=h_{21}^d=\mathrm{e}^{-\mathrm{i} \pi /4}$, and 
requiring the terms involving $m_{h'}$, $m_{s'}$, and $m_{s_2}$
separately to contribute
to $\epsilon_K$ an amount less than the experimental value
of that 
quantity ($|\epsilon_K|=2.26 \times 10^{-3}$ \cite{Groom:in}) give the limits

\begin{equation}
\label{epsilonK}
\left( \frac{v}{u} \right)^2 \frac{\sin^2 \phi}{m_{h'}^2},\; \left( \frac{v}{u} \right)^2 \frac{\cos^2 \phi}{m_{s'}^2} \leq 
\frac{3.9 \times 10^{-6}}{{\rm GeV}^2}, \;\;\;\;\; 
\left( \frac{v}{u} \right)^2 \frac{1}{m_{s_2}^2} \leq 
\frac{3.8 \times 10^{-5}}{{\rm GeV}^2}.
\end{equation}

If we take $m_{h'} \simeq 10^2$ GeV, as suggested by experiment, then
Eq. (\ref{epsilonK}) implies that $(v/u) \sin \phi \leq \frac{1}{5}$, which is not
a very stringent bound. Substituting this into Eq. (\ref{edmlimit}), one sees that
the electron edm can easily be near the present published experimental limit.
For instance, taking $(v/u) \sin \phi \simeq 0.1$, so that flavon exchange 
contributes of order 1/5 of the experimental value of $\epsilon_K$,
and taking $m_{s'} \simeq 300$ GeV, Eq. (14) gives $d_e \sim (0.6
\times 10^{-27} \;e\,{\rm cm}) \cos \phi $. 

Comparing Eqs. (\ref{de}) and (\ref{muegamma}) (in which the unknown parameters, 
$\sin \phi$, $m_{s'}$, and $u$, enter in exactly the same way)
reveals that the present limits on the decay $\mu \rightarrow 
e + \gamma$ and the electron edm are
about equally sensitive to flavon exchange in this model. 
For example, if the
CP-violating phases are large and all $h_{ij}$ are close to one,
as was assumed in deriving Eqs. (\ref{edmlimit}) and (\ref{muegamma}), and $d_e$ is just
below the present limit, then the rate for $\mu \rightarrow e + 
\gamma$ is roughly a forth of the present limit.  

One sees here the importance of the fact that the diagonal Yukawa
couplings of the flavon field $s_1$ have phases suppressed by
$\epsilon^2 \simeq 2 \times 10^{-2}$. Were it not so, then
the present limit on the electron edm would imply that the rate for
$\mu \rightarrow e + \gamma$ was at least four orders of magnitude below
present limits (unless parameters were fine-tuned).
$\mu + N \rightarrow e + N$ would also be suppressed.

Turning to $\mu$-$e$ conversion on nuclei, one sees from 
Eq. (\ref{mueconversion}), that the present limit on
this is also, for a wide range of parameters,  
about as sensitive to flavon-exchange as are the present limits on
$d_e$ and $\mu \rightarrow e + \gamma$. For example, if 
$(v/u) \sin \phi \simeq \frac{1}{5}$, then the first term on the
left-hand side of Eq. (\ref{mueconversion}) (i.e. the term proportional to
$\sin \phi \cos \phi/m_{h_0}^2$) gives a contribution to the rate
for $\mu$-$e$ conversion that is about an order of magnitude below
the present limit. 
However, in some regions of parameter space, $\mu + N \rightarrow e + N$ can be
the most sensitive to flavon exchange. Suppose, for example, that
$v/u$ is smaller, but not much smaller, than one, and that 
$\sin \phi \ll 1$. Then both $d_e$ and $\mu \rightarrow e + \gamma$
are highly suppressed, whereas $\mu + N \rightarrow e + N$ need not
be because of the term that goes as $\frac{\cos^2 \phi}{m^2_{s'}}$ on the left-hand side
of Eq. (\ref{mueconversion}). 

\begin{figure} 
\begin{center} 
\includegraphics[width=4in]{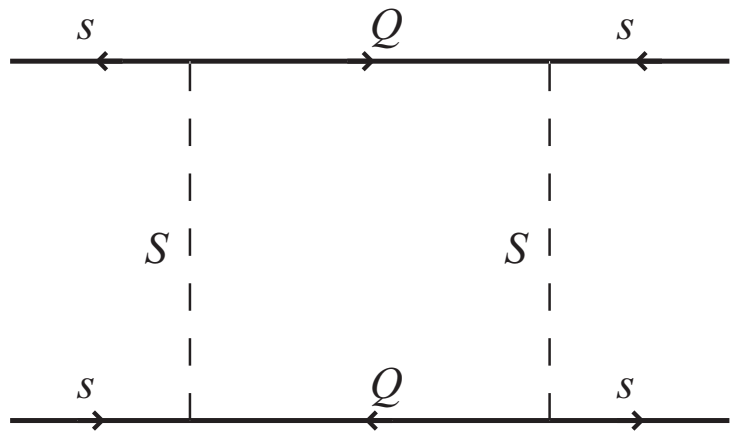} 
\caption{\label{KKbar1} Box diagram contribution to $K-\overline{K}$ mixing. 
The internal fermion $Q$ has mass of order $M_F$.} 
\end{center} 
\end{figure}

We have only considered the effects arising from the effective Yukawa 
terms in Eq. (1). However, there is another source of flavor violation
from flavon exchange that can be very important. To get the effective
low energy Yukawa terms in Eq. (1), fermions having mass of order $M_F$
are integrated out. There are diagrams involving these heavy fermions that
can be important. The most important such diagram is that shown in 
Fig. \ref{KKbar1}, which is a contribution to $K-\overline{K}$ mixing. The internal
fermion has mass of order $M_F$. The external fermion is the $s_0$ quark,
i.e. the $s$ quark in the original basis in which the Yukawa matrices 
of Eq. (2) are written. When one goes to the physical basis of the light 
quarks, $s_0$ will contain a small admixture of the physical $d$ quark:
$s_0 = s + O(\epsilon^2) d$. Consequently, there will be from Fig. \ref{KKbar1}
a $\Delta S = 2$ piece that goes as $\epsilon^4$. The Yukawa couplings
in Fig. \ref{KKbar1} may be assumed to be of order unity. (The only reason the effective
Yukawa couplings of the known light quarks are small is that they are
suppressed by powers of $\epsilon^2$, since they arise from integrating
out heavy fermions. However, in the underlying theory
containing those heavy fermions there is no reason for the
Yukawa couplings to be small.) The coefficient of the $\Delta S = 2$
operator arising from Fig. \ref{KKbar1} should therefore typically
be of order $(16 \pi^2)^{-1}
\epsilon^4 (1/M_F^2) = (16 \pi^2)^{-1} \epsilon^8 u^{-2}$. Using
$\epsilon^2 \sim 2 \times 10^{-2}$ and $u \sim 300$ GeV, one has that
the coefficient of the $\Delta S = 2$ term is of order $10^{-14}$ GeV$^{-2}$.
With some of the phases or couplings being assumed somewhat smaller than one,
the contribution from Fig. \ref{KKbar1} can easily be within the limit set by 
$\epsilon_K$.

\section{Conclusions}

We presented a simple flavon model that can accommodate the observed 
hierarchy of the charged fermion masses and mixings in terms of the powers 
of one small parameter. It has been shown that the flavor-diagonal couplings 
of the flavon field, under a general set of assumptions, are real to the leading 
order in that parameter. This implies that flavor changing and CP violating 
signatures, $d_e$, $\mu \rightarrow e + \gamma$, and $\mu$-$e$ conversion on 
nuclei, can be equally near the present experimental limits with all other low energy
constraints satisfied. For significant range of parameters $\mu$-$e$ conversion 
can be the most sensitive place to look for such signatures.

\end{document}